\documentclass[12pt]{article}

\usepackage{psfig}

\newlength{\figwidth}
\newcounter{eqabc}
\newcommand{\eqabcA}[2] { \refstepcounter{equation} 
\begin{center}
 $ \displaystyle #1 $\\[-#2] \makebox[\textwidth][r]{(\theequation a)}
\end{center} 
}

\newcommand{\eqabcB}[3] { \addtocounter{equation}{-1} \refstepcounter{equation} 
\begin{center}
 $ \displaystyle #1 $\\[-#2] \makebox[\textwidth][r]{(\theequation #3)}
\end{center} 
}

\newcommand{\eabe} {\begin{eqnarray}}
\newcommand{\eaen} {\end{eqnarray}}
\newcommand{\eqbe} {\begin{equation}}
\newcommand{\eqen} {\end{equation}}
\newcommand{\bibl}[5]
	{#1, {\it #2} {\bf #3} (#4) #5}

\newcommand{\ee} {$e^+e^-$~}

\newcommand{\anti}[1] {${ \ol \mrm #1 }$}
\newcommand{\pair}[1] {${\mrm {#1 \ol #1} }$}
\newcommand{\mrm} {\mathrm}
\newcommand{\srm}[1] {_{\mathrm{#1}}}
\newcommand{\N}[1] {N\srm{#1}}
\newcommand{\Nqq} {{N\srm{q\ol q}}}
\newcommand{\Ngg} {{N\srm{gg}}}

\newcommand{\YY} {{Y\srm q + Y\srm{\ol q}}}
\newcommand{\SP} {\mrm{Le}}
\newcommand{\Lund} {\mrm{Lu}}
\newcommand{\Le} {\mrm{Le}}

\newcommand{\ol} {\overline}
\newcommand{\Ordo}{{\cal O}}

\newcommand{\CF} {{C\srm F}}

\newcommand{\Nc} {{\N c}}

\begin{document}
\begin{titlepage}
\begin{flushright}
LU TP 99-07\\
June 1999
\end{flushright}
\vspace{25mm}
\begin{center}
  \Large
  {\bf On Particle Multiplicities in Three-jet Events} \\
  \vspace{12mm}
  \normalsize
  Patrik Ed\'en\footnote{Department of Theoretical Physics, Lund University, Helgonav. 5, S-223 62 Lund, Sweden},\\
  G\"osta Gustafson\footnotemark[\thefootnote],\\
  Valery Khoze\footnote{INFN-Laboratori Nazionali di Frascati, P.O. Box 13, I-00044, Frascati (Roma), Italy}

\end{center}
\vspace{2cm} 
{\bf Abstract:} \\ 
A thorough verification of the distinct differences in the properties
of quark and gluon jets is considered as one of the most instructive
tests of the basic ideas of QCD. In the real life experiments such a
comparison appears to be quite a delicate task and various subtle
issues require further theoretical efforts.  In this paper we discuss
in detail the possibility to extract the theoretically adequate
information from the particle multiplicity patterns in three-jet
events in $e^+e^-$ annihilation.
\end{titlepage}
\section{Introduction}\label{sec:intro}

As well known, the larger colour charge of gluons ($C_A=\Nc=3$)
compared to quarks ($\CF=(\Nc^2-1)/2\Nc=4/3$) leads to various
distinctive differences between the two types of jets, for recent articles  see e.g.~\cite{jetscales} and the review~\cite{khozeochs}. Thus, a detailed comparison
of the properties of quark and gluon jets provides one of the most
instructive tests of the basic ideas of QCD. An experimental
verification of these differences has been a subject of quite
intensive investigations, especially in the last years,
e.g.~\cite{hamacher}. However, obtaining of the
theoretically adequate information about the properties of the gluon
jet appears to be not an easy task. Recall that the analytical QCD
results address the comparison between the energetic gluon and quark
jets emerging from the point-like colourless sources, and that (unlike
the \pair q case) the pure high energy gg events at present are not
available experimentally.\footnote{In principle, it is possible to
create a pure source of the colour singlet gg events at a future
linear \ee collider through the process $\gamma\gamma\rightarrow
\mrm{gg}$~\protect{\cite{khozegg}}.}

So far, most studies of the structure of gluon jets have been
performed in three-jet events of \ee annihilation. As a rule, these
rely on a jet finding procedure both for selection of the \pair qg
events and for a separation between the jets in an event. Without
special care, such an analysis is inherently ambiguous and may suffer
from the lack of the direct correspondence to the underlying
theory. Recently some more sophisticated approaches have been
exploited (see e.g.~[3, 5-8]) which allow better theoretical
significance. There are still a number of issues which are frequently
overlooked in the present gluon jet analyses and some further
theoretical efforts are required. First of all, this concerns particle
multiplicity distributions in the jets. Clarification of these issues
is the main aim of this paper. More detailed description of the
theoretical framework can be found in ref.~\cite{jetscales}.

In particular the following problems are addressed.

1. Different approaches to the three-jet studies employ different
definitions of the \pair qg kinematics. In particular, this concerns
such a key variable as a transverse momentum scale of the gluon,
$p_\perp$.  Our first issue here is to discuss an exact definition of
this quantity, which governs radiation from the gluon.

2. The definition of the three-jet topology with the gluon registered
at a given $p_\perp$ imposes an obvious requirement that there are no
other subjets in the event with the transverse momentum exceeding
$p_\perp$. We have to investigate quantitatively the impact of this
requirement on the jet sample.

3. To calculate predictions from perturbative QCD, using the
assumption of local parton hadron duality (LPHD)~\cite{LPHD}, a cutoff
is needed for the infrared singularities. As discussed in detail in ref.~\cite{jetscales} such a cutoff depends on the
soft hadronization process and can {\em not} be uniquely specified
from perturbative QCD alone. Thus, the result is necessarily model
dependent.

In what follows we discuss these three issues successively in sections
2, 3 and 4, and in section 5 we study their effect on analyses of
3-jet events in $e^+e^-$-annihilation.

\section{Definition of $p_\perp$}\label{sec:ptdef}
In the simplest case of soft radiation, $p_\perp$ can be easily
defined, as the quark and antiquark specify a unique direction.  For
large $p_\perp$ gluons, however, the q and \anti q get recoils such
that there is no obvious direction against which the transverse
momentum should be measured. To have well defined expressions such a
direction has to be specified. In the Lund dipole formalism~[10-16] 
$p_\perp$ has been defined according to (subscript $\Lund$ for Lund) 
\eqbe
p_{\perp \Lund}^2 \equiv \frac{s\srm{qg}s\srm{g\ol q}}{s}, \label{e:ptLdef} 
\eqen 
where $s\srm{qg}$ denotes the squared
mass of the quark-gluon system etc. In this particular frame the
gluon rapidity is given by the expression 
\eqbe
y=\frac{1}{2}\ln(\frac{s\srm{qg}}{s\srm{g\ol q}}).\label{e:yLdef}
\eqen 
The kinematically allowed region is given by 
\eqbe 
p_{\perp\Lund} < \frac{\sqrt s}2;~~|y| <\ln\left(\frac{\sqrt s}{p_{\perp\Lund}}\right) \equiv \frac{1}{2} (L -{\kappa_\Lund});~~~L\equiv\ln(\frac{s}{\Lambda^2}),~\kappa_\Lund \equiv \ln(\frac{p_{\perp\Lund}^2}{\Lambda^2}).  \label{e:yptranges} 
\eqen 
These variables have
the advantage that the phase space element usually expressed in the
scaled energy variables $ x\srm q$ and $ x\srm{\ol q}$ is exactly
given by the simple relation 
\eqbe s \mrm dx\srm q\mrm dx\srm{\ol q} = \mrm dp_{\perp \Lund}^2\mrm dy.\label{e:PSelement}
\eqen 
As discussed in section~\ref{sec:qqg}, $p_{\perp\Lund}$ may also work
well as a scale parameter in the QCD cascade.

An alternative definition has also been used in the literature, e.g.\ by the Leningrad group~\cite{ADKT,khoze}
\eqbe p_{\perp \SP}^2 \equiv \frac{s\srm{qg}s\srm{g\ol q}}{s\srm{q\ol q}}. \label{e:ptSPdef} \eqen
This definition corresponds to the gluon transverse momentum in the $\mrm{q\ol q}$ cms (with respect to the $\mrm{q\ol q}$ direction). It is notable that in this frame the gluon rapidity is also exactly given by the expression in Eq~(\ref{e:yLdef}). The two $p_\perp$-definitions agree for soft gluons, but deviate for harder gluons. While $p_{\perp \Lund}$ is always bounded by $\sqrt s/2$, $p_{\perp \SP}$ has no kinematic upper limit in the massless case. 

\section{Bias from restrictions on subjet transverse momenta}\label{sec:bias}

The effect of a cutoff in $p_\perp$ has been discussed previously~\cite{Lundbias,Catanibias}. Here we give a brief review of the results, in order to end the section with an investigation of the numerical importance of subleading terms. These are essential for a correct analysis of three-jet events, which will be discussed in section~\ref{sec:qqg}.

To see the qualitative features of the bias we first study \ee$\rightarrow$ \pair q events within the Leading Log approximation (LLA). The quark and antiquark emit gluons according to the well-known radiation pattern
\eqbe \mrm dn\srm g \approx \CF\frac{\alpha_s}{\pi} \frac{\mrm dx\srm q \mrm dx\srm{\ol q}}{(1-x\srm q)(1-x\srm{\ol q})} = \CF\frac{\alpha_s(p_\perp^2)}{\pi} \frac{\mrm d
p_\perp^2}{p_\perp^2} \mrm dy \equiv \CF \frac{\alpha_s(\kappa)}{\pi} \mrm d\kappa
\mrm dy;~~~~\kappa \equiv \ln(p_\perp^2/\Lambda^2). \label{e:dng} \eqen
We have here used Eq~(\ref{e:PSelement}), and in the following we define $p_\perp$ and $y$ according to Eqs~(\ref{e:ptLdef}) and~(\ref{e:yLdef}), unless otherwise stated.

Due to colour coherence the hadronic multiplicity $\N g(\kappa)$ in a
gluon jet depends on the $p_\perp$ of the gluon and not on its energy (see, e.g., refs~\cite{ADKT,khoze}).
Summing up the contributions from all gluons in a cascade we arrive at the
average multiplicity $\Nqq(L=\ln(s/\Lambda^2))$ in the original $\mrm{q\ol q}$ system~[13, 15-18]
(Refs~[15-18]
include also nonleading terms.)
\eqbe
\Nqq(L) \approx \int_{\kappa_0}^{L} \mrm d\kappa \int_{-\frac{1}{2} (L -
\kappa)}^{\frac{1}{2} (L -
\kappa)} \mrm dy \CF \frac{\alpha_s(\kappa)}{\pi} \N g(\kappa) =
\int_{\kappa_0}^{L} \mrm d\kappa (L - \kappa)\CF \frac{\alpha_s(\kappa)}{\pi}
\N g(\kappa). \label{e:Nq}  \eqen
(We have here introduced a lower cutoff $\kappa_0$ for the integral over transverse momentum. This point will be discussed in section~\ref{sec:cutoff}.)
Taking the derivative with respect to $L$ we find
\eqbe 
N^\prime\srm {q\ol q}(L) \approx \int_{\kappa_0}^{L} \mrm d\kappa \CF \frac{\alpha_s(\kappa)}{\pi}
\N g(\kappa). \label{e:Nqprime} \eqen

Consider now a sample of events selected in such a way that there are {\em no} subjets with $p_\perp>p_{\perp\mrm{cut}}$. (Within a $k_\perp$-based cluster scheme with a resolution parameter $p_{\perp\mrm{cut}}$, this means that there are only two primary q and \anti q jets.)
To obtain the multiplicity $\Nqq(L,\kappa\srm{cut})$ in this biased sample, we
must restrict the $\kappa$ integral in Eq~(\ref{e:Nq}) to the region $\kappa < \kappa\srm{cut}$. We then find~\cite{Lundbias}
\eabe
\Nqq(L,\kappa\srm{cut}) & \approx & 
\Nqq(\kappa\srm{cut}) +
(L - \kappa\srm{cut}) N^\prime\srm{q\ol q}(\kappa\srm{cut}).
\label{e:Nqtwoscale} \eaen
The first term corresponds to two cones around the q and \anti q jet directions. Here the $p_\perp$ of the emissions is limited by the kinematical constraint in Eq~(\ref{e:yptranges}) rather than by $\kappa\srm{cut}$. It also corresponds exactly to an unbiased $\mrm{q\ol q} $ system with cms energy $p_{\perp\mrm{cut}} $. The second term describes a central rapidity plateau of width $(L - \kappa\srm{cut})$, in which the limit for gluon emission is given by the constraint $\kappa\srm{cut}$. This expression for a two-jet event can be generalized for a biased multi-jet configuration, and a similar discussion applies also to the multiplicity variance, cf.\ ref~\cite{Lundbias}. (Similar equations for biased two-jet and three-jet events were later discussed also in ref~\cite{Catanibias}.)

The average particle multiplicity in the selected two-jet sample is smaller than in an unbiased sample. The modification due to the bias is similar to the suppression from a Sudakov form factor. It is formally $\Ordo(\alpha_s)$, but it also contains a factor $\ln^2(s/p_{\perp}^2)$. Thus, it is small for large $p_\perp$-values but it becomes significant for smaller $p_\perp$. This clearly demonstrates that the multiplicity in this restricted case depends on {\em two} scales, $\sqrt s$ and $p_{\perp\mrm{cut}}$. The $p_\perp$ of an emitted gluon is related to the virtual mass of the radiating parent quark. Therefore, the two scales $\sqrt s/2$ and $p_{\perp\mrm{cut}}$ represent the energy and virtuality of the quark and antiquark initiating the jets. 

\begin{figure}[tb]
\begin{center}
	\mbox{\psfig{figure=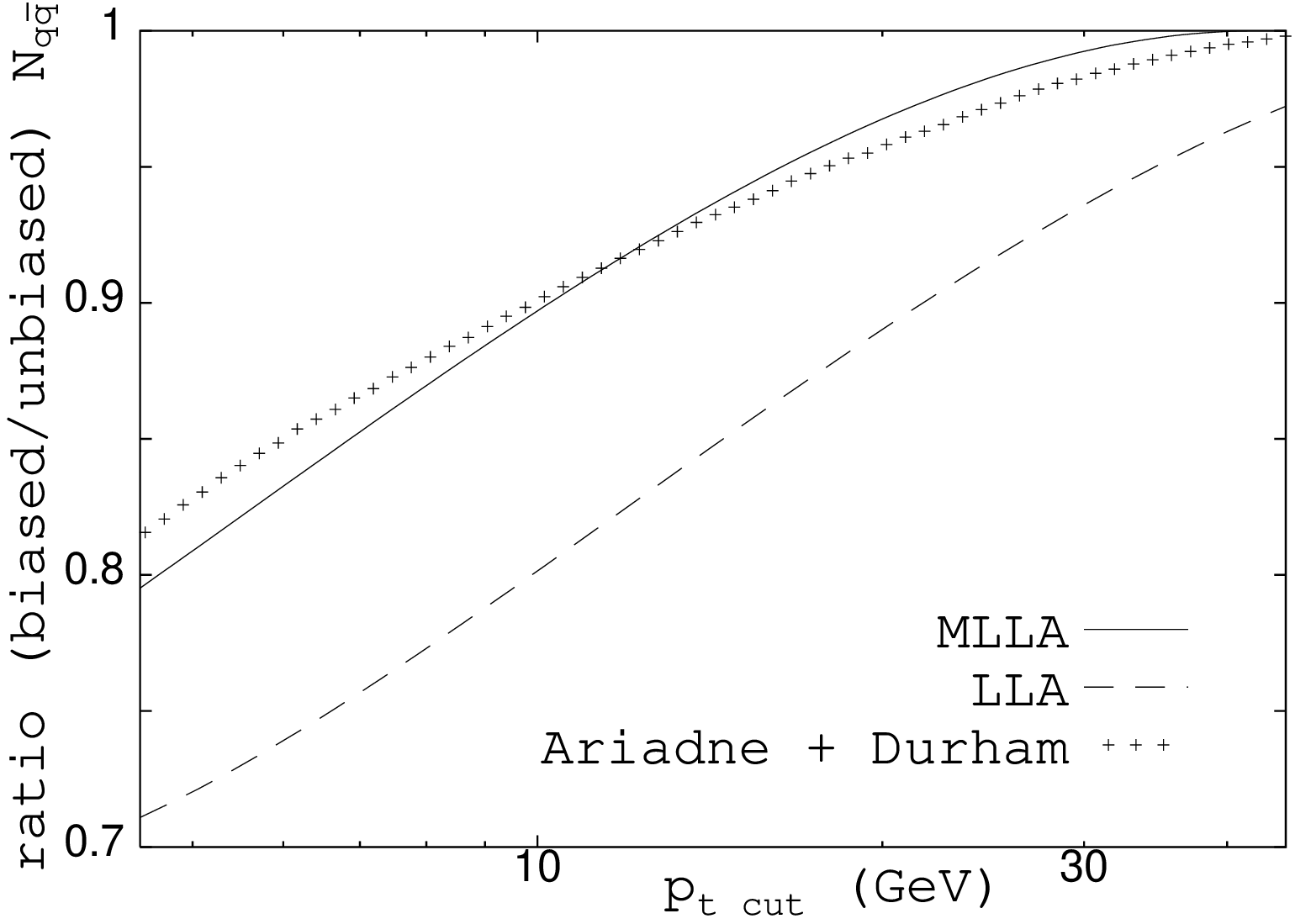,width=0.5\figwidth}}
\end{center}
  \caption{
{\em  The effect from the bias due to a constraint $p_{\perp\mrm{cut}}$ on emitted subjets, at 90GeV energy. The figure shows the ratio of biased over unbiased multiplicities  as a function of $p_{\perp\mrm{cut}}$. The results for LLA and MLLA relations (Eqs~(\ref{e:Nqtwoscale}) and~(\ref{e:NqtwoscaleMLLA}), respectively) differ significantly from each other. The result of the MLLA relation in Eq~(\ref{e:NqtwoscaleMLLA}), using the  $p\srm{\perp}$ definition in Eq~(\ref{e:ptLdef}), is in good agreement with \textsc{Ariadne} MC and Durham cluster algorithm results.}
}
  \label{f:twoscale}
\end{figure}
Though the LLA result in Eq~(\ref{e:Nqtwoscale}) describes the qualitative features of the bias, subleading corrections are needed for a quantitative analysis. Within  the Modified Leading Log approximation (MLLA)~\cite{MLLA}, subleading terms are included, which affect the prediction for the unbiased multiplicities and, thus, implicitly also the biased multiplicity in Eq~(\ref{e:Nqtwoscale}). Furthermore, it is in~\cite{jetscales} shown that the expression in Eq~(\ref{e:Nqtwoscale}) for the biased multiplicity is explicitly changed when MLLA corrections are considered.
An unbiased system should be restored when $p\srm{\perp cut}$ approaches the kinematical limit $\sqrt s/2$, but the r.h.s.\ of Eq~(\ref{e:Nqtwoscale}) equals the unbiased quantity $\Nqq(L)$ only when $p\srm{\perp cut} = \sqrt s$. The relation consistent with the MLLA is~\cite{jetscales}
\eabe
\Nqq(L,\kappa\srm{cut}) & \approx & 
\Nqq(\kappa\srm{cut}+c\srm q) +
(L - \kappa\srm{cut}-c\srm q) N^\prime\srm{q\ol q}(\kappa\srm{cut}+c\srm q);~~~c\srm q = \frac32.
\label{e:NqtwoscaleMLLA} \eaen

The bias is illustrated in Fig~\ref{f:twoscale}. The dotted line shows results from the \textsc{Ariadne} MC~\cite{ariadne}, when the Durham cluster algorithm~\cite{Durham} is used to define a biased sample of events classified as two-jet events with a $y\srm{cut}$ equal to $p^2_{\perp\mrm{cut}}/s$. The MC results agree well with the prediction of Eq~(\ref{e:NqtwoscaleMLLA}), where for $p_{\perp\mrm{cut}}$ we have used the $p_\perp$-definition in Eq~(\ref{e:ptLdef}) (solid line). The predicted effect is below 5\% for  $p_{\perp\mrm{cut}} > 20$GeV, but increases rapidly for smaller $p_{\perp\mrm{cut}}$. 

Fig~\ref{f:twoscale} presents also the result using the LLA  relation in Eq~(\ref{e:Nqtwoscale}) (dashed line). To elucidate the effect of the differences between Eq~(\ref{e:Nqtwoscale}) and~(\ref{e:NqtwoscaleMLLA}), we have used the same expression for the unbiased quantities $\Nqq$ and  $N^\prime\srm{q\ol q}$. (These are obtained by a simple fit to \textsc{Ariadne} MC results, which are in good agreement with the MLLA\@.) As seen, the subleading terms are important; the LLA  relation significantly overestimates the effect.
To our knowledge experimental data for this bias have not been presented. Such data should be obtainable in a rather straightforward analysis, which thus readily could test the accuracy of the MC result or the MLLA relation.

\section{Infrared cutoffs}\label{sec:cutoff}
Gluon radiation diverges for collinear and soft emissions. Therefore, to
estimate the hadronic multiplicity from the assumption of LPHD~\cite{LPHD}, a cutoff is needed. 
Naturally, the cutoff must be Lorentz invariant. 
For collinear emissions a single Feynman
diagram dominates, and there are two possibilities, the virtual mass,
$\mu$, of the emitting parent parton or the transverse momentum, $p_\perp$,
of the emitted gluon measured relative to the parent parton direction. 
These quantities are connected by the relation
\eqbe p_\perp^2 = \mu^2 z(1-z), \eqen
where $z$ equals the light cone momentum fraction taken by the emitted gluon. 
The transverse momentum is directly related to the
formation time, and, therefore, we regard this as the most natural choice
for a cutoff. (For a further discussion see ref~\cite{jetscales}.)

For soft emissions no obvious cutoff is available, however. As several
Feynman diagrams contribute and interfere, there is no unique parent
parton. Consequently $\mu^2$ or $p_\perp^2$ cannot be uniquely specified and,
therefore, cannot be directly used. (Obviously a cut in energy is not
possible, as this is not Lorentz invariant.)

For soft emissions from a single $\mrm{q\ol q}$ colour dipole a cutoff in $p_\perp$
is still the natural choice if measured in the cms, where the $\mrm q$ and
$\mrm{\ol q}$ move back to back. For emissions from a more complicated state the
situation simplifies greatly in the large-$\N c$ limit, as many
interference
terms disappear. In this limit the emission corresponds to a set of {\em
independent} colour dipoles~\cite{dip,earlyLund}. 
The natural choice for the cutoff is then
$p_\perp$ in the cms of the emitting dipole (measured with respect to the
dipole direction). We note that this implies that the soft gluons
connect the hard partons in exactly the same way as the string in the
string fragmentation model~\cite{string}, which illustrates the connection between perturbative QCD and the string model~\cite{ADKT}.

For the physical case with 3 colours, extra interference terms appear
with relative magnitude $1/\N c^2$~\cite{ADKT,DKT...}. Here nonplanar
Feynman diagrams contribute, and it is impossible to uniquely specify
a parent parton or a relevant $p_\perp$.  Thus, a more fundamental
understanding of confinement is needed to specify the cutoff, which
cannot be determined from perturbative QCD alone~\cite{jetscales}.  In
hadronization models the $1/\N c^2$ interference terms correspond to
the problem of ``colour reconnection'', and different models have been
proposed~\cite{reconnection}. None of these can be motivated from
first principles, and only experimental data can differentiate among
the various models.

In spite of the formal uncertainties, the success of current Monte
Carlo programs~\cite{MC,ariadne} indicate that the colour suppressed
interference terms do not have a very large effect. This is also
supported by recent searches by OPAL of the reconnection effects in
hadronic $Z$ events~\cite{OPAL}. In most parton cascade formalisms,
a cascade cutoff motivated in the large-$\Nc$ limit is used also for
finite $\Nc$. The colour interference effects are accounted for by
reducing the colour factor from $\Nc/2$ to $\CF$ in regions collinear
with quarks and antiquarks, and, due to colour coherence, also in some
parts of the central rapidity region. We note, however, that some
subtle interference phenomena, as a matter of principle, cannot be
absorbed into a probabilistic scheme, see~\cite{DKT...} for
details. These are still awaiting a thorough experimental test.

\section{Formalism for three-jet events}\label{sec:qqg}
After these general discussions we are now ready to consider three-jet
\pair qg systems. To simplify the discussion we first study the
large-$\Nc$ limit. The emission of softer gluons from a
$\mrm{q\ol{q}g}$ system corresponds then to two dipoles which emit gluons independently.  If a gluon jet is resolved
with transverse momentum $p_\perp$, this imposes a constraint on the
emission of subjets from the two dipoles. Thus, the contribution from
each dipole is determined by an expression like
Eq~(\ref{e:NqtwoscaleMLLA}).  For relatively soft primary gluons the
constraint should be given by $p_{\perp\mrm{cut}} =
p_{\perp\mrm{g}}$. For hard gluons $p_{\perp\Lund}$ is of
the same order as its parent quark virtuality, and in ref~\cite{conny}
it is shown that $\Ordo(\alpha_s^2)$ matrix elements are well described if $p_{\perp\Lund}$ is used as an ordering
parameter for the perturbative cascade. This is also indicated by the successful applications of the \textsc{Ariadne} MC\@. We will, therefore, assume that the constraint on further
emissions is well described by the identification $p_{\perp\mrm{cut}}
= p_{\perp\Lund}$. The multiplicity in a qg dipole
with an upper limit on $p_\perp$ can, just as for the \pair q case discussed in
section~\ref{sec:bias}, be described as two forward jet regions and a
central plateau.  

We note that if the three-jet events were selected using a
cluster algorithm with a {\em fixed} resolution scale
, then the constraint on subjet transverse momenta, $p_{\perp\mrm{cut}}$, would be 
smaller than the $p_\perp$ of the
gluon jet (as the gluon jet was resolved). In this case most jet definitions give three jets which are all biased~\cite{Lundbias,Catanibias}. We will, however, here focus
on three-jet configurations obtained by iterative clustering until
exactly three jets remain, {\em without} a specified resolution scale, where hence the constraint on subjet $p_\perp$ is 
described by $p_{\perp\mrm{cut}} = p_{\perp\Lund}$. As we will see, this implies that the bias on the gluon jet is negligible, which makes this selection procedure suitable for an investigation of unbiased gluon jets.

For finite $\Nc$ the different dipoles in a multi-parton configuration can not be completely independent of each other. However, encouraged by the success of MC programs, let us assume that the main effect of finite $\Nc$ is that the colour factor, which determines
softer gluon emission, is reduced from $\Nc/2$ to $\CF$ in the domains
where the emission is dominated by radiation from the quark or the
antiquark leg. Let us assume that a rapidity range $Y\srm q$ in the $\mrm{qg}$
dipole is similar to a corresponding range in a $\mrm{q\ol q}$ dipole,
while the remaining range $L\srm{qg} - Y\srm q$ is similar to a range in
one half of a $\mrm{gg}$ system. The corresponding ranges in the
$\mrm{g\ol q}$ dipole are $Y\srm{\ol q}$ and $L\srm{g\ol q} -
Y\srm{\ol q}$. This implies that the total multiplicity in the \pair qg event corresponds to the expression 
\eqbe \N{q\ol qg} = \Nqq(\YY,\kappa_{\Lund}) + \frac1 2\Ngg(L\srm{qg} + L\srm{g\ol q} - Y\srm{ q}-Y\srm{\ol q},\kappa_{\Lund}). \label{e:3scales} \eqen
For the constraint $p_{\perp\mrm{cut}}$ we have here written $\kappa_\Lund$, which is appropriate for the selection procedure discussed above.

As discussed in section 4, the size of  $Y\srm{q}$ and $Y\srm{\ol q}$ cannot be uniquely determined within perturbative QCD. Possibly the most natural choice is to assume that the quantity  $\YY$ corresponds to the energy in the \pair q subsystem~\cite{khoze}, which implies
\eqabcA {\YY \approx \ln(s\srm{q\ol q}/\Lambda^2) \equiv L\srm{q\ol q}. \label{e:YqqSP}}{3ex}
The relation in Eq~(\ref{e:YqqSP}a) can be regarded as an educated guess, but a finite shift cannot be excluded. In ref~\cite{GG} it is assumed that 
\eqabcB {\YY \approx \ln (s/\Lambda^2) = L, \label{e:YqqL}}{3ex}b
which agrees with Eq~(\ref{e:YqqSP}a) to leading order. For relatively soft gluons we have $s\srm{q\ol q} \approx s$, and in this case Eqs~(\ref{e:YqqSP}a) and~(\ref{e:YqqL}b) are approximately equivalent. The assumption in Eq~(\ref{e:YqqSP}a)  implies that the energy scale for the gluon term is given by $L\srm{qg} + L\srm{g\ol q} - L\srm{q\ol q} = \kappa_{\SP}$. Similarly we get from Eq~(\ref{e:YqqL}b) the corresponding gluonic energy scale $\kappa_{\srm{Lu}}$.

The effect of the $p_\perp$ constraint is rather different in the two terms in Eq~(\ref{e:3scales}). For the gluon term the energy scale is in general only slightly larger than the bias scale $\kappa_\Lund$. This implies that in most cases the bias can be disregarded in this term. Inserting the different assumptions in Eqs~(\ref{e:YqqSP}a) and~(\ref{e:YqqL}b)
into Eq~(\ref{e:3scales}) then gives
\eqabcA {\N{q\ol qg}  \approx  \N {q\ol q}(L\srm{q\ol q},\kappa_\Lund) + \frac{1}{2}\N {gg}( \kappa_{\SP}), \label{e:Nqqg}}{3.3ex}
\vspace{-4ex}
\eqabcB{\N{q\ol qg}  \approx \Nqq(L,\kappa_\Lund) + \frac1 2\Ngg(\kappa_\Lund).\label{e:NqqgL}}{3.3ex}b
We note that the consistency between Eqs~(\ref{e:Nqqg}a) and~(\ref{e:NqqgL}b) follows from the fact that the total rapidity range in the two dipoles, $L\srm{qg} + L\srm{g\ol q}$, can be expressed in two different ways by the equalities $L\srm{qg} + L\srm{g\ol q} = L\srm{q\overline q} + \kappa_\Le = L+ \kappa_\Lund $. In particular, we see from these equalities that the argument in $\Ngg$ has to be $p_{\perp\SP}^2$ in Eq~(\ref{e:Nqqg}a) and $p_{\perp\Lund}^2$ in Eq~(\ref{e:NqqgL}b), and not e.g.\ $(2p_\perp)^2$.

\begin{figure}[tb]
\parbox{0.47\textwidth}{
 \mbox{\psfig{figure=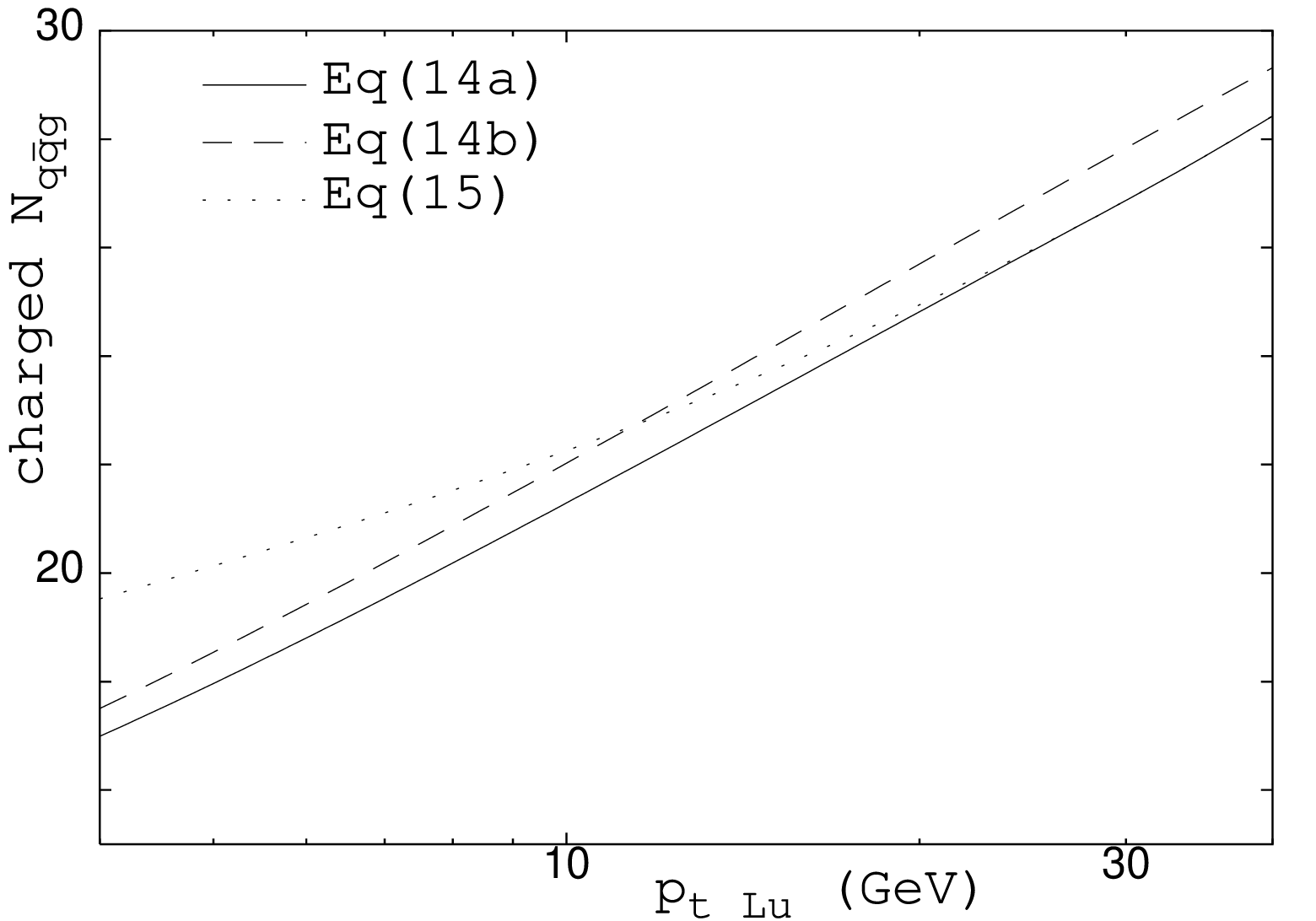,width=0.45\textwidth}}
\caption{ {\em $\N{q\ol qg}$ as a function of $p_{\perp\Lund}$ for $\sqrt{s\srm{q\ol q}} = $ 60 GeV. The different predictions from Eqs~(\ref{e:Nqqg}a,b) and~(\ref{e:NqqgSP}) illustrates the importance of the bias at moderate $p_\perp$.}}
\vspace{6ex}
  \label{f:Nqqg}
}
\hspace{0.03\textwidth}
\parbox{0.47\textwidth}{
 \mbox{\psfig{figure=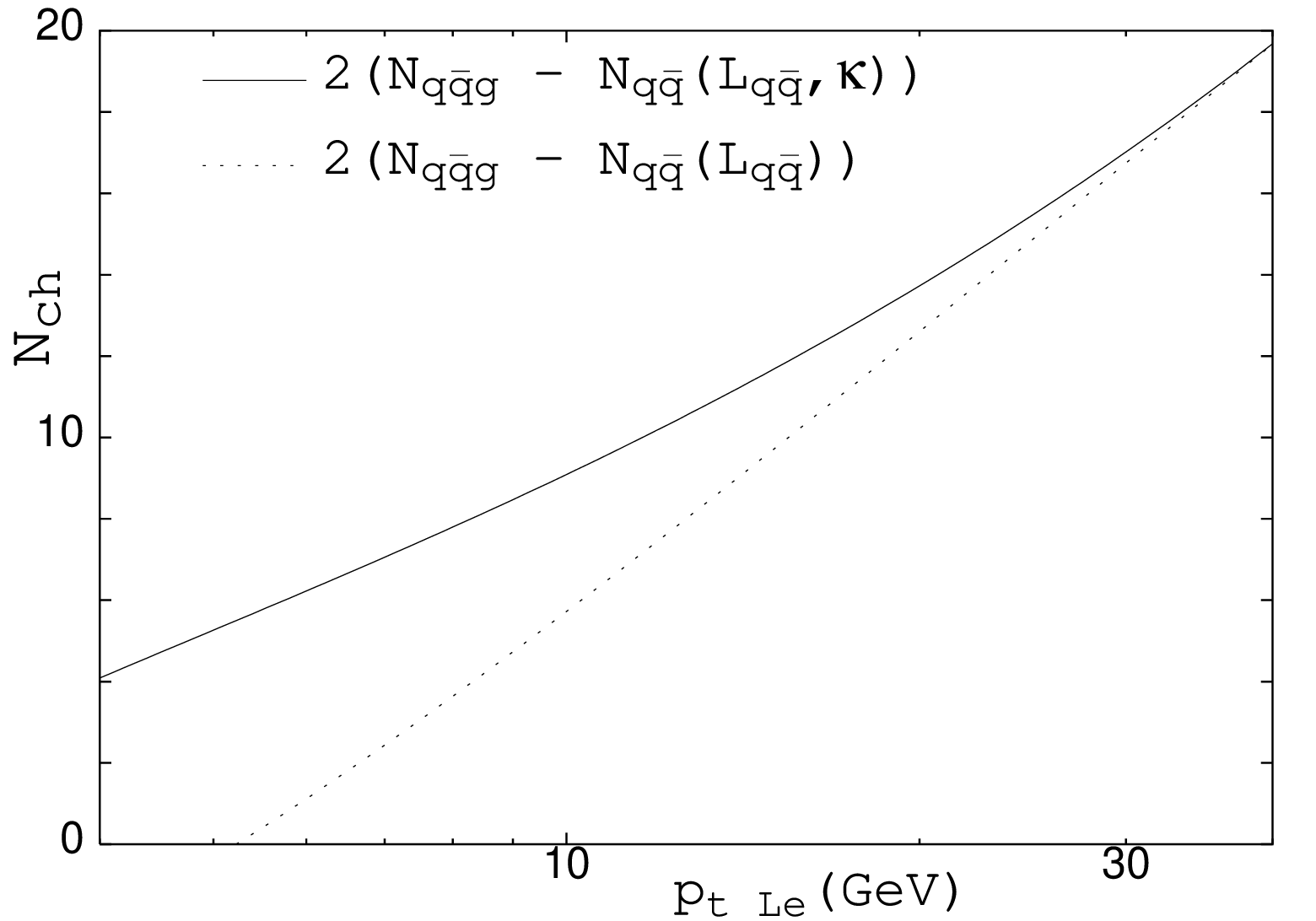,width=0.45\textwidth}}
\caption{{\em The prediction for $\Ngg$, obtained by subtracting from $\N{q\ol qg}$ the quark contribution $\Nqq$, changes significantly if the bias in the \pair q-term is neglected. The figure shows the effect for $\sqrt{s\srm{q\ol q}}=60$GeV, with $\N{q\ol qg}$ given by Eq~(\ref{e:Nqqg}a).}}
  \label{f:Ngg}
}
\end{figure}

The leading effect of a finite shift in $ \YY$ is colour-suppressed, and therefore not expected to be large. However, subleading corrections introduce a difference between the results of Eqs~(\ref{e:Nqqg}a)  and~(\ref{e:NqqgL}b). This is seen in 
Fig~\ref{f:Nqqg}, where the difference is approximately 1 particle for $\sqrt{s\srm{q\ol q}}=60$GeV.
In the calculations of $\N{q\ol qg}$ in Fig~\ref{f:Nqqg}, we have used the expressions in~\cite{jetscales} for the multiplicities $\Nqq$ and $\Ngg$. These include MLLA corrections and recoil effects, which implies that $\Ngg < 2\Nqq$ for accessible energies. Consequently, the result for $\N{q\ol qg}$ grows with the assumed value of $\YY$. 

While the bias is not serious for the gluon term in
Eq~(\ref{e:3scales}), it is more important for the \pair q term. Focusing
on events with comparatively large values of $p_{\perp }$, where
the bias is less essential, and using the assumption in
Eq~(\ref{e:YqqSP}a), we arrive at the result of ref~\cite{khoze}: 
\eqbe \N{q\ol qg}(s,p_{\perp \SP}^2) = [\N{q\ol q}(s\srm{q\ol q})+ \frac1 2\Ngg(p_{\perp \SP}^2)](1+ \Ordo(\alpha_s)).\label{e:NqqgSP} \eqen
 The bias is formally of order $\alpha_s$, and is here taken into account by the factor $(1+\Ordo (\alpha_s))$. The result of this expression, neglecting the $\Ordo (\alpha_s)$ term, is also shown in Fig~\ref{f:Nqqg}. The effect of the bias corresponds to less than one charged particle for $p_{\perp \mrm{cut}}$ larger than $\sim 10$GeV, but becomes much more important for smaller   $p_{\perp \mrm{cut}}$-values. 

An alternative way to express this result is the effect on extracting $\N{gg}$ from data for $\N{q\ol q g}$, as illustrated in Fig~\ref{f:Ngg}. $\Ngg$ can be extracted by subtracting the biased quark multiplicity $\Nqq(L\srm{q\ol q},\kappa_\Lund)$ from $\N{q\ol qg}$, here assumed to be described by Eq~(\ref{e:Nqqg}a). Neglecting the bias in the subtracted $\Nqq$ term gives a significantly different result. The relative effect of the bias is in this case larger, and it exceeds 20\% for $p_\perp < 15$GeV. Furthermore, to get a reliable result for $\Ngg$, the relevance of subleading terms in the biased quark multiplicity needs to well understood. For the solid line in Fig~\ref{f:Ngg}, the MLLA relation in Eq~(\ref{e:NqtwoscaleMLLA}) is used to subtract the \pair q contribution from the total multiplicity. Instead using the LLA relation in Eq~(\ref{e:Nqtwoscale}) would give a prediction for $\Ngg$ which is about three charged particles higher for most values of $p\srm{\perp cut}$.

Although the effect of the bias is very important for small $p_\perp$, we also see from Figs~\ref{f:Nqqg} and~\ref{f:Ngg} that it can be neglected for large $p_\perp$-values, where, thus, the results in ref~\cite{khoze} and Eq~(\ref{e:NqqgSP}) can be safely used. This implies e.g.\ that the bias is negligible in gluon systems defined as the hemisphere opposite to two quasi-collinear quark jets, thoroughly investigated by OPAL~\cite{JWGary,OPAL}.

It would be very interesting to compare the results in Figs~\ref{f:Nqqg}  and~\ref{f:Ngg} to experiments. Experimental data on $\N{q\ol qg}$ can be directly compared to the Monte Carlo or MLLA results in Fig~\ref{f:Nqqg}, Data on the difference $\N{q\ol qg} - \Nqq$ can be compared either to the predictions in Fig~\ref{f:Ngg} or to experimental results for $\Ngg$ obtained through one of the methods described in ref~\cite{jetscales}. We have compared the results in Fig~\ref{f:Nqqg} with MC simulations, where the $p_\perp$ scale is determined by the Durham cluster algorithm. The MC results (not shown) indicate that an analysis based on  jet reconstruction is accurate enough to illustrate the effects the bias, but perhaps not to distinguish between the assumptions in Eq~(\ref{e:Nqqg}a) and~(\ref{e:NqqgL}b). We also note that the effects described here may have a phenomenological impact on the recent analysis of $\N{q\ol qg}$~\cite{DELPHI}, which employs the two-scale dependence.

\section{Conclusion}
A series of subtle effects influence an analysis of the difference
between quark and gluon jets in a real life experiment. In this letter
we discuss and clarify effects associated with
\begin{itemize}
\item the definition of $p_\perp$,
\item the bias from restrictions on subjet $p_\perp$,
\item the problem that infrared cutoffs cannot be uniquely defined from perturbative QCD.
\end{itemize}
We also demonstrate the impact of these effects on the analysis of
three-jet events in $e^+e^-$-annihilation.

\subsubsection*{Acknowledgments}
We thank K. Hamacher, W. Ochs, R. Orava and T. Sj\"ostrand for useful discussions.

This work was supported in part by the EU Fourth Framework Programme `Training and Mobility of Researchers',
Network `Quantum Chromodynamics and the Deep Structure of Elementary Particles', contract FMRX-CT98-0194
(DG 12 - MIHT).

\end{document}